\documentclass[twocolumn]{aastex631}
\usepackage{CJK}
\usepackage{amsmath}
\usepackage{comment}

\newcommand{\msfont}{
  \bfseries
  \color{magenta}
}

\begin{document}
\begin{CJK*}{UTF8}{gbsn}
\title{A Computational Model for Ion and Electron Energization during Macroscale Magnetic Reconnection}

\author[0009-0009-7643-6103]{Zhiyu Yin (尹志宇)}
\affiliation{Department of Physics, University of Maryland, College Park, MD 20742, USA}
\affiliation{IREAP, University of Maryland, College Park, MD 20742, USA}

\author[0000-0002-9150-1841]{J. F. Drake}
\affiliation{Department of Physics, University of Maryland, College Park, MD 20742, USA}
\affiliation{IREAP, University of Maryland, College Park, MD 20742, USA}
\affiliation{Institute for Physical Science and Technology, University of Maryland, College Park, MD, 20742, USA}
\affiliation{Joint Space Science Institute, University of Maryland, College Park, MD, 20742, USA}

\author[0000-0002-5435-3544]{M. Swisdak}
\affiliation{IREAP, University of Maryland, College Park, MD 20742, USA}
\affiliation{Joint Space Science Institute, University of Maryland, College Park, MD, 20742, USA}
\begin{abstract}
A set of equations are developed that extend the macroscale magnetic reconnection simulation model \textit{kglobal} to include particle ions. The extension from earlier versions of \textit{kglobal}, which included only particle electrons, requires the inclusion of the inertia of particle ions in the fluid momentum equation. The new equations will facilitate the exploration of the simultaneous non-thermal energization of ions and electrons during magnetic reconnection in macroscale systems. 
Numerical tests of the propagation of Alfv\'en waves and the growth of firehose modes in a plasma with anisotropic electron and ion pressure are presented to benchmark the new model.  

\end{abstract}


\section{Introduction} \label{sec:intro}

Magnetic reconnection is a fundamental process that occurs in a diverse array of astrophysical and laboratory plasmas \citep{Tsuneta96,Yamada97,Lin03,Gosling05,Phan06,Burch16,Gary18}. It allows for the rapid reconfiguration of the magnetic field, facilitating the release of magnetic energy into kinetic, thermal energy and non-thermal particle energization \citep{Parker57,Petschek64}. An important consequence of magnetic reconnection is the efficient acceleration of charged particles into power-law (i.e., non-thermal) distributions that extend many decades in energy \citep{Lin03,Krucker10,Oka13,Oieroset02,Ergun20}. Such acceleration events are of significant interest, not only because of their scientific relevance in astrophysical phenomena such as solar flares and magnetospheric substorms  but also because of the implications for controlled fusion experiments \citep{Yamada94} and space weather \citep{Angelopoulos20}.

Many computational simulations, most notably those performed with particle-in-cell (PIC) and magnetohydrodynamic (MHD) \citep{Biskamp86,Karpen12,Dahlin22} models, have limitations that compromise their utility for studying non-thermal particle energization in such systems. PIC simulations correctly model kinetic physics by resolving kinetic scales \citep{Shay98a,Hesse99,Shay07,Daughton11,Li19,zhang2021} but the spatial scales associated with solar flares can exceed $10^4$ km, while the Debye length---a measure of the scale of kinetic effects---can be of the order of centimeters, a scale separation of $10^{10}$. Hence, computational costs drastically limit the ability of PIC simulations to simulate the extensive range of scales occurring in magnetic reconnection events. A consequence of this inadequate separation of scales is the demagnetization of energetic electrons and ions in PIC simulations of reconnection, which in turn inhibits particle energy gain and prevents the formation of the extended power laws that are observed in nature \citep{Li19,zhang2021}. 

Conversely, MHD models employ a fluid approximation and can simulate large-scale phenomena. While capable of representing macroscopic events, MHD simulations frequently assume the plasma is thermal (i.e., represented by a single temperature) and hence by definition cannot produce non-thermal particles. The electric and magnetic fields from MHD models have been used to explore non-thermal energization of test particles.  However, these test particles do not interact self-consistently with the fields so energy is not conserved, which limits the test particle model's fidelity \citep{Onofri06}.

The original \textit{kglobal} model combined aspects of the PIC and MHD descriptions to address electron acceleration in large-scale astrophysical reconnection \citep{Drake19,Arnold19}. Particle electrons are distributed on the MHD grid, as in PIC models, but are evolved with MHD fields in the guiding-center approximation in which all kinetic scales, including the Larmor radii of particles are ordered out of the equations. Notably, \textit{kglobal} includes the effects of Fermi reflection within evolving magnetic flux ropes and the pressure anisotropy that feeds back on the MHD fields to reduce the tension force that drives magnetic reconnection. Magnetic reconnection simulations with \textit{kglobal} produced, for the first time, extended power-law spectra of non-thermal electrons that aligned with existing observational data \citep{Arnold21}.

However, magnetic reconnection can simultaneously produce non-thermal electrons and non-thermal protons \citep{Lin03,Emslie12}.  Hence, developing a self-consistent simulation model is vital to elucidate the mechanisms of ion and electron energization during magnetic reconnection and the resulting partitioning of energy between the two species. In this paper we extend the \textit{kglobal} equations to include particle ions, which requires the inclusion of the inertia of the particle ions in the bulk fluid ion momentum equation. The resulting equations will facilitate the exploration of the self-consistent energization of both ions and electrons during magnetic reconnection. In Section \ref{sec:model} we introduce the equations that describe the model. In Section \ref{sec:test} we present test results for Alfv\'en wave propagation in a system with a finite pressure anisotropy and the firehose instability. In Section \ref{sec:conclusion} we present an overview of the resulting equations and implications for exploring particle energization during reconnection in macroscale systems.

\section{Model Equations} \label{sec:model}
For completeness, we first outline the governing equations for the original \textit{kglobal} model before turning to the additions made in the upgraded version.

\subsection{ \textit{kglobal} with particle electrons }\label{subsec:kglobal}

The original model includes three distinct plasma species: the ion fluid (number density $n_{i}$), the electron fluid ($n_{ef}$), and particle electrons ($n_{ep}$) \citep{Drake19,Arnold19}. No conversion between species is allowed. The particle electrons are governed by the guiding-center equations given in \cite{Northrop63}.  Their perpendicular velocity is determined by conservation of the magnetic moment:
\begin{equation}\label{eq:ep_perp}
    \mu_{ep} = \frac{p_{ep,\perp}^2}{2B} = \text{const.}
\end{equation}
where $p_{ep}$ is the momentum and $B$ is the magnitude of the local magnetic field. Their perpendicular motion is given by the usual MHD ${\bf E}\times {\bf B}$ drift while they stream parallel to the ambient magnetic field at their parallel velocity. 
The variation in the parallel velocity reflects contributions from Fermi reflection, magnetic mirroring, and the large-scale parallel electric field
\begin{equation}\label{eq:ep_par}
\frac{d p_{ep,\parallel}}{dt} = p_{ep,\parallel}\mathbf{v}_E\boldsymbol{\cdot \kappa} - \frac{\mu_e}{\gamma_e}\mathbf{b}\boldsymbol{\cdot\nabla}B - eE_{\parallel}   
\end{equation}
with $\mathbf{v}_E$ the $\mathbf{E}\boldsymbol{\times}\mathbf{B}$ drift velocity, $\mathbf{b} = \mathbf{B}/B$ a unit vector in the direction of the magnetic field,  $\boldsymbol{\kappa} = \mathbf{b}\boldsymbol{\cdot\nabla}\mathbf{b}$ the field curvature, and $\gamma_e$ the relativistic Lorentz factor.  The guiding-center electrons move through the MHD grid and, crucially, do not form a thermal distribution. Their gyrotropic pressure tensor feeds back on the MHD momentum equation. 

The ion fluid forms an MHD-like backbone that satisfies the usual continuity equation
\begin{equation}\label{eq:if_den}
\frac{\partial n_i}{\partial t} = -\boldsymbol{\nabla \cdot}(n_i\mathbf{v}_i)
\end{equation}
and a modified momentum equation 
\begin{equation}\label{eq:if_mom}
\begin{split}
m_i\frac{\partial (n_i\mathbf{v}_i)}{\partial t} = &-\boldsymbol{\nabla\cdot}\mathbb{T}_i -(\boldsymbol{\nabla\cdot}\mathbb{T}_{ef})_\perp+ \mathbf{J}\boldsymbol{\times}\mathbf{B}/c  \\ &- (\boldsymbol{\nabla\cdot}\mathbb{T}_{ep})_{\perp} + en_iE_{\parallel}\mathbf{b},
\end{split}
\end{equation}
where the ion stress tensor with inertial contributions is
\begin{equation}\label{eq:ion_tensor}
\mathbb{T}_{i}  = P_{i}\mathbb{I}+m_in_{i}\mathbf{v}_i\mathbf{v}_i,
\end{equation}
with $P_i$ the ion scalar pressure and $\mathbb{I}$ the unit tensor. The corresponding electron fluid tensor is 
\begin{equation}\label{eq:ef_tens}
\mathbb{T}_{ef}  = P_{ef}\mathbb{I}+m_en_{ef}\gamma_{ef}v^2_{ef,\|}\mathbf{b}\mathbf{b},
\end{equation}
with $P_{ef}$ the electron fluid scalar pressure. Finally,
$\mathbb{T}_{ep}$ is  the particle electron gyrotropic stress tensor,
including inertial contributions,
\begin{equation}\label{eq:ep_tens}
\mathbb{T}_{ep}  = T_{ep\parallel} \mathbf{b}\mathbf{b} + P_{ep\perp} (\mathbb{I} - \mathbf{b}\mathbf{b}),
\end{equation}
where \( T_{ep\parallel} \) is the component of the stress tensor along the magnetic
field \( B \), and \( P_{ep\perp} \) is the usual perpendicular pressure,

\begin{equation}\label{eq:ep_press_perp}
P_{ep\perp} = \int d\mathbf{p}_e \frac{p^2_{e\perp}}{2m_e\gamma_e} f,
\end{equation}
where in the frame drifting with \( \mathbf{v}_E = c\mathbf{E} \times \mathbf{B} / B^2 \) there are no perpendicular flows so \( f = f(x, p_{e\parallel}, p_{e\perp}, t) \). \( T_{ep\parallel} \) includes the mean parallel drifts of the particle electrons,

\begin{equation}\label{eq:ep_press_par}
T_{ep\parallel} = \int d\mathbf{p}_e \frac{p^2_{e\parallel}}{m_e\gamma_e} f,
\end{equation}
with \( p_{e\parallel} \) being the electron particle parallel momentum. 

Note that in the ion momentum equation the electron fluid and particle inertia are neglected compared to the ion inertia on the left side of Eq.~(\ref{eq:if_mom}), while the electron fluid and particle pressures are retained on the right side of this equation. This result is based on an ordering in which the ion velocity scales as the Alfv\'en speed and the electron particle and fluid thermal velocities and parallel streaming velocities are of the order of the electron Alfv\'en speed, $C_{Ae}=\sqrt{m_i/m_e}C_A$, where the Alfv\'en speed is defined as \(C_A = \sqrt{B^2 / (4\pi m_i n_{i})}\). In this ordering the electron momentum scales as $m_eC_{Ae}=\sqrt{m_e/m_i}m_iC_A$ and so is much smaller than that of the ion fluid. In contrast, the electron pressure is $m_eC_{Ae}^2=m_iC_A^2$, which is the same as the ion pressure.Thus, the pressure terms on the right side of Eq.~(\ref{eq:if_mom}) must be retained even though the electron inertia in this equation is neglected.  Finally, the ions follow an adiabatic equation of state
\begin{equation}\label{eq:if_press}
\frac{d}{dt}\left(\frac{P_i}{n_i^{\gamma}}\right) = 0
\end{equation}
where the adiabatic index $\gamma$ is taken to be $5/3$.

The number density of the electron fluid is given by
\begin{equation}\label{eq:quasi_neutrality}
n_{ef}=n_i - n_{e p}
\end{equation}
while the parallel motion follows from the requirement of vanishing parallel current
\begin{equation}
    n_{ef}v_{ef,\parallel} = n_iv_{i,\parallel} - n_{ep}v_{ep,\parallel}
    \label{eq:current_constraint}
\end{equation}
The zero current condition follows from the constraint on the current from Amp\`ere's law
\begin{equation}
   \frac{V_d}{C_A}=\frac{J}{neC_A} \sim \frac{Bc}{4\pi neC_AL}\sim \frac{d_i}{L}\ll 1,
\end{equation}
where $V_d$ is the drift speed associated with the current, $d_i$ is the ion inertial scale length and $L$ is the characteristic magnetic scale length. In the {\it kglobal} model all kinetic scales are ordered out, which yields the zero current condition in Eq.~(\ref{eq:current_constraint}). 
The fluid electrons also follow an adiabatic equation of state
\begin{equation}\label{eq:ef_press}
\frac{d}{dt}\left(\frac{P_{ef}}{n_{ef}^{\gamma}}\right) = 0
\end{equation}

The magnetic field is advanced via Faraday's law,
\begin{equation}\label{eq:b}
    \frac{\partial \mathbf{B}}{\partial t} = -c\boldsymbol{\nabla\times}\mathbf{E}_{\perp},
\end{equation}
with 
\begin{equation}\label{eq:eperp}
\mathbf{E}_{\perp} = -\frac{1}{c}\mathbf{v}_i \boldsymbol{\times}\mathbf{B}    
\end{equation}
Finally, $E_{\parallel}$ is produced by magnetic-field-aligned gradients in the electron pressure and is given by \citep{Arnold19} 
\begin{equation}\label{equ:e_par}
E_{\|} = \\ -\frac{1}{e(n_{ep}+n_{ef})} \left( \mathbf{b} \boldsymbol{\cdot \nabla \cdot} \mathbb{T}_{ep} + \mathbf{b} \boldsymbol{\cdot \nabla} \mathbb{T}_{ef} \right).
\end{equation}
Thus, Equations \eqref{eq:ep_perp}-\eqref{equ:e_par} constitute the complete set of equations for the {\it kglobal} model with particle electrons. 
\subsection{\textit{kglobal} with particle electrons and ions}

In the upgraded version of \textit{kglobal}, particle ions are introduced as a new species and, as is the case for the particle electrons, are treated in the guiding-center limit. Similar to the original model, the populations of the four plasma species do not change after initialization. The basic ordering of the electrons and the equations describing their dynamics are basically unchanged from the original model. Minor modifications to the charge neutrality and zero current conditions are discussed below, but the details of the electron dynamics that were discussed in earlier papers are not repeated here. 

Due to the small mass of the electrons, their inertial terms in the fluid ion momentum equation were neglected in the original \textit{kglobal} equations.  That simplification is not possible for the particle ions in the upgraded model, which complicates the derivation and the final form. However, the perpendicular motion of the fluid and particle ions is still governed by the same ${\bf E}\times {\bf B}$ drift. As a consequence, the generic form of the ion momentum equation, can be written as 
\begin{equation} \label{equ:mom_tot}
\frac{\partial}{\partial t}m_i(n_{if}{\bf v}_{if}+n_{ip}{\bf v}_{ip})={\bf S}
\end{equation} 
with ${\bf v}_{if}$ and $n_{if}$ the fluid ion velocity and density, ${\bf v}_{ip}$ and $n_{ip}$ the particle ion velocity and density,  and ${\bf S}$ the usual momentum driver, including convective inertial terms. The differences in the inertia between the fluid and particle ions arises solely from the parallel motion so the total momentum equation takes the form 
\begin{equation}
    \frac{\partial}{\partial t}m_in_i{\bf v}_{if}+\frac{\partial}{\partial t}m_in_{ip}(v_{ip\parallel}-v_{if\parallel}){\bf b}={\bf S}.
\end{equation}
where $n_i = n_{if} + n_{ip}$.  The inertial terms parallel to {\bf B} are easily evaluated, which leaves an equation for the fluid ion momentum that includes the inertia of the particle ions. The details of the evaluation of the parallel inertial terms are found in the Appendix. 

Most of the other equations in Section \ref{subsec:kglobal} remain unchanged, aside from minor notation adjustments (e.g., $n_i \rightarrow n_{if}+n_{ip}$).  Specifically, Equations \eqref{eq:ep_perp} and \eqref{eq:ep_par} governing the motion of the particle electrons are unchanged. Equations \eqref{eq:if_den} and \eqref{eq:if_press} become equations for the ion fluid density, 
\begin{equation}\label{eq:if_den_new}
\frac{\partial n_{if}}{\partial t} = -\boldsymbol{\nabla \cdot}(n_{if}\mathbf{v}_{if})
\end{equation}
 and ion fluid pressure, 
\begin{equation}\label{eq:if_press_new}
\frac{d}{dt}\left(\frac{P_{if}}{n_{if}^{\gamma}}\right) = 0.
\end{equation}
The guiding center equations for the particle ions are essentially identical to those of the electrons in Equations (\ref{eq:ep_perp}) and (\ref{eq:ep_par}),
\begin{equation}\label{eq:ip_perp}
    \mu_{ip} = \frac{p_{ip,\perp}^2}{2B} = \text{const.},
\end{equation}
and 
\begin{equation}\label{eq:ip_par}
\frac{d p_{ip,\parallel}}{dt} = p_{ip,\parallel}\mathbf{v}_E\boldsymbol{\cdot \kappa} - \mu_i\mathbf{b}\boldsymbol{\cdot\nabla}B + eE_{\parallel}   
\end{equation}
The ions are taken to be non-relativistic. 
Including the new species also requires changes in the expressions for the electron fluid density
\begin{equation}\label{eq:charge_neutrality_new}
n_{ef} = n_{if} + n_{ip} -n_{e p}
\end{equation}
and parallel motion
\begin{equation}\label{eq:ef_current_new}
    n_{ef}v_{ef,\parallel} = n_{if}v_{if,\parallel} + n_{ip}v_{ip,\parallel}- n_{ep}v_{ep,\parallel}
\end{equation}
The adiabatic equation of state for the electron fluid, Equation \eqref{eq:ef_press}, remains the same as do equations \eqref{eq:b}, and \eqref{eq:eperp} for $\mathbf{B}$ and $\mathbf{E}_{\perp}$.  Equation \eqref{equ:e_par} for $E_{\parallel}$ remains the same because only the electrons contribute.

In both the original and upgraded model, electron inertia can be discarded compared with that of the ions in the ion momentum equation, as discussed in Sec.~\ref{subsec:kglobal}. The distinction between the inertia of collective and individual particle electrons, along with their movements in relation to the ambient magnetic field and their specific interactions with the MHD fluid, is comprehensively discussed in \cite{Drake19}. 
However, as shown in equation (\ref{equ:mom_tot}), the inertia of the particle ions cannot be neglected. The derivation of the new fluid ion momentum equation is detailed in Appendix \ref{sec:ap_der}, with the final result being 
\begin{equation}\label{eq:if_mom2}
\begin{split}
m_i&\frac{\partial\, (n_{i}\mathbf{v}_{if})}{\partial t} =- \boldsymbol{\nabla \cdot}\mathbb{T}_{if} -(\boldsymbol{\nabla \cdot}\mathbb{T}_{ef})_\perp +
\mathbf{J}\boldsymbol{\times}\mathbf{B}/c  \\& -
\boldsymbol{\nabla\cdot}\mathbb{T}_{ip}- (\boldsymbol{\nabla\cdot}\mathbb{T}_{ep})_{\perp}+en_iE_{\parallel}\mathbf{b}-\mathbf{I}_{i\parallel},
\end{split}
\end{equation}
where $\mathbf{I}_{i\parallel}$ is the difference in the parallel inertia between the fluid and particle ions, given by 
\begin{equation}
\begin{aligned}\label{eq:if_mom_par}
\mathbf{I}_{i\parallel}&=\frac{\partial}{\partial t}m_in_{ip}(v_{ip\parallel}-v_{if\parallel})\mathbf{b}=\\& -\mathbf{b}\,m_iv_{if\parallel}\bigg(\frac{n_{ip}}{n_{if}}\boldsymbol{\nabla\cdot}n_{if}
v_{if\parallel}\mathbf{b}- \\&\hspace{20mm} \boldsymbol{\nabla\cdot}n_{ip}
v_{ip\parallel}\mathbf{b}-n_{if}\mathbf{v}_{i,\perp}\boldsymbol{\cdot\nabla}
\frac{n_{ip}}{n_{if}}\bigg) \\&+\frac{n_{ip}}{n_{if}}\mathbf{b}\mathbf{b}\boldsymbol{\cdot\nabla\cdot}\mathbb{T}_{if}- \mathbf{b}\mathbf{b}\boldsymbol{\cdot\nabla\cdot}\mathbb{T}_{ip}\\&+ m_in_{ip}\frac{v_{ip\parallel}-v_{if\parallel}}{B} (\mathbb{I}-\mathbf{b}\mathbf{b}) \boldsymbol{\cdot} \frac{\partial \mathbf{B}}{\partial t}
\end{aligned}
\end{equation}
where $n_i = n_{ip}+n_{if}$, and the ion fluid stress tensor, $\mathbb{T}_{if}$,  is defined in Equation (\ref{eq:ion_tensor}). The particle ion pressure tensor is given by
\begin{equation}\label{eq:ip_tens}
\mathbb{T}_{ip}  = P_{ip\parallel} \mathbf{b}\mathbf{b} + P_{ip\perp} (\mathbb{I} - \mathbf{b}\mathbf{b}) +m_in_{ip}\mathbf{v}_{ip}\mathbf{v}_{ip},
\end{equation}
with the parallel and perpendicular pressures calculated in the frame of the bulk motion of the particle ions,
\begin{equation}\label{eq:ip_press_perp}
P_{ip\perp} = \int d\mathbf{p}_i \frac{p^2_{i\perp}}{2m_i} f,
\end{equation}
and 
\begin{equation}\label{eq:ip_press_par}
P_{ip\|} = \int d\mathbf{p}_i \frac{p^2_{i\|}}{m_i} f
\end{equation}
with $p_{ip}$ the ion momentum. Note that in the inertia term on the left side of Equation (\ref{eq:if_mom2}), $n_i$ is the total ion density and not just the fluid ion density $n_{if}$. In the limit $n_{ip} \rightarrow 0$, $\mathbf{I}_{i\parallel} \rightarrow 0$ and $\mathbb{T}_{ip}\rightarrow 0$ and equation \eqref{eq:if_mom2} reduces to equation \eqref{eq:if_mom}. 
The ratios $n_{ip}/n_{if}$ and $n_{ep}/n_{ef}$ must be chosen at $t=0$, but are free to evolve in space and time during simulations.  The electron spectra reported in previous {\it kglobal} reconnection simulations were insensitive to $n_{ep}/n_{ef}$ \citep{Arnold21}, but the sensitivity of reconnection simulations with both particle electrons and ions to these ratios will also need to be explored.

To summarize, the {\it kglobal} model with particle electrons and ions consists of Eqs.~(\ref{eq:ep_perp})-(\ref{eq:ep_par}), (\ref{eq:ef_tens})-(\ref{eq:ep_press_par}), (\ref{eq:ef_press})-(\ref{equ:e_par}), and (\ref{eq:if_den_new})-(\ref{eq:ip_press_par}).

\section{Alfv\'en Wave Propagation and growth of firehose modes in an Anisotropic Plasma} \label{sec:test}
Particle energy gain in reconnection is dominantly parallel to the ambient magnetic field, with the implication that the parallel pressure exceeds the perpendicular pressure in the heated plasma as reconnection develops. As a consequence, the cores of magnetic islands approach the marginal firehose condition at late time in simulations. The tension in a bent magnetic field line, which is the fundamental driver of reconnection, goes to zero at the marginal firehose condition. Thus, the pressure anisotropy that develops as reconnection proceeds regulates reconnection and particle energy gain \citep{Drake06a,Drake12,Drake19,Arnold19}. Therefore, to ensure that the updated \textit{kglobal} model properly reproduces the influence of pressure anisotropy on the magnetic field dynamics, we benchmark the model with a circularly polarized Alfv\'en wave mode propagating along a uniform magnetic field in a system with pressure anisotropy ($P_{\parallel} \neq P_{\perp}$), a problem with a well-established  solution \citep{Parker57}. With increased pressure anisotropy, the reduction in magnetic tension of a propagating Alfv\'en wave reduces the wave phase speed. In a second benchmarking test on the dynamics of the pressure anisotropy of both particle ions and electrons, we explore the growth of the firehose instability.  

The upgraded \textit{kglobal} model builds upon the original \textit{kglobal} framework, the details of which can be found in the references \citep{Drake19, Arnold19}. The code is written in normalized units. A reference magnetic field strength \(B_0\) and the initial total ion density \(n_{i0} = n_{ip} + n_{if}\) define the Alfv\'en speed, \(C_A = \sqrt{B_0^2 / (4\pi m_i n_{i0})}\). Lengths and times are normalized to an arbitrary macroscale length, \( L \), and the Alfv\'en crossing time, \( \tau_A = L / C_A \), respectively.  Electric fields and temperatures are normalized to \( C_AB_0/c \) and \( m_iC_A^2 \), respectively.  The \textit{kglobal} model employs second-order diffusivities and fourth-order hyperviscosities ($\propto \nu \nabla^4 $) in each fluid equation to mitigate noise at the grid scale \citep{Drake19}.

The simulations were performed within a two-dimensional (2D) spatial domain, coupled with a three-dimensional velocity space, featuring an initially uniform magnetic field of magnitude $B_0$ oriented along the $x$-axis. The domain measures $2\pi L \times \pi L$ with \( 512 \times 256 \) cells with periodic boundary conditions on all sides. Each grid cell is initialized with 200 self-consistent (i.e., not test) particles, 100 ions and 100 electrons. The initial density ratios everywhere are \(n_{ip}/n_{if} = n_{ep}/n_{ef} = 1/3\). The ion-to-electron mass ratio is set to 25, although the results of the simulations are insensitive to this value. The scalar temperature of both fluid species is set to $0.1$. For the particle ions and electrons, the perpendicular temperature is initially set to $0.1$ while the parallel temperature is adjusted to control the magnitude of the anisotropy. 

We introduce perturbations to the magnetic field and velocity in the form of a circularly polarized Alfv\'en wave, with a wavelength equal to the length of the simulation box ($k_x = 2\pi/L$). During a propagation interval of \( 3\tau_A \), we measured the wave's velocity in order to compare it with the theoretical phase speed \( C_p \) of an Alfv\'en wave, which is given by
\begin{equation}
    C_p = C_A \sqrt{1 - \frac{4\pi(P_{\parallel} - P_{\perp})}{B^2}} \equiv \alpha C_A
\end{equation}
where \( \alpha = \sqrt{1 - 4\pi(P_{\parallel} - P_{\perp})/B^2} \), aligns with the Chew-Goldberger-Low (CGL) framework in the linear regime where pressure perturbations are negligible \citep{Chew56}. Here, \( P_{\parallel} \) and \( P_{\perp} \) denote the combined ion and electron pressures along and perpendicular to the magnetic field, respectively.

\begin{figure}
\centering
\includegraphics[width=\columnwidth]{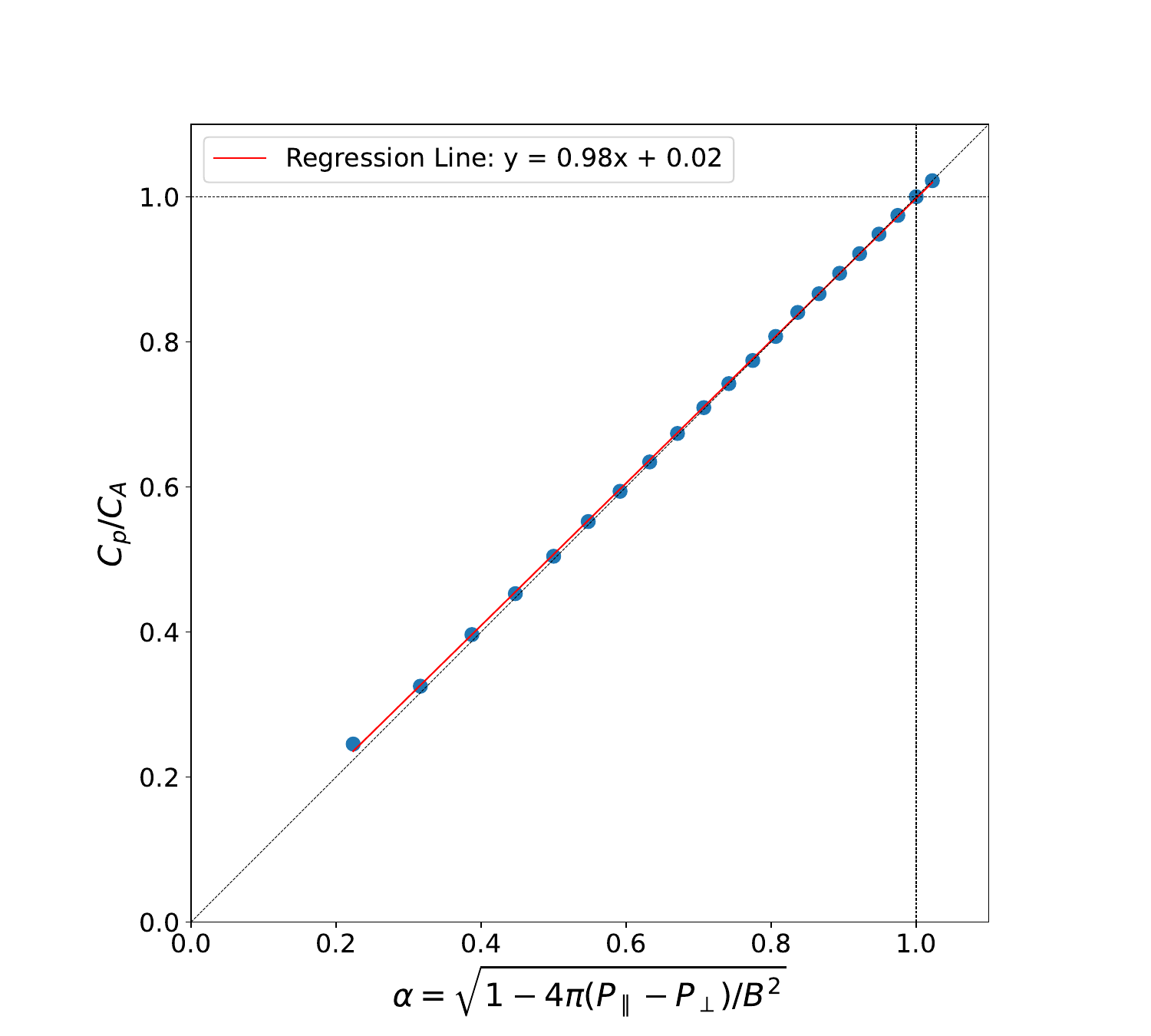}
\caption{The observed phase speed of the Alfv\'en wave, \( C_p \), plotted against the anisotropy parameter \( \alpha \). The blue dots are the results of simulations. The intersection of the vertical and horizontal dotted lines marks the isotropic Alfv\'en wave.
\label{fig:linear}}
\end{figure}

The results are shown in Fig \ref{fig:linear}, where \( C_p \), normalized to \( C_A \) is plotted against \( \alpha \).  There is excellent agreement between the code results and linear wave theory, highlighting the precision of the \textit{kglobal} model and suggesting that the impact of ion and electron pressure anisotropy on reconnection dynamics will be properly included in the new model. 

\begin{figure}
\centering
\includegraphics[width=\columnwidth]{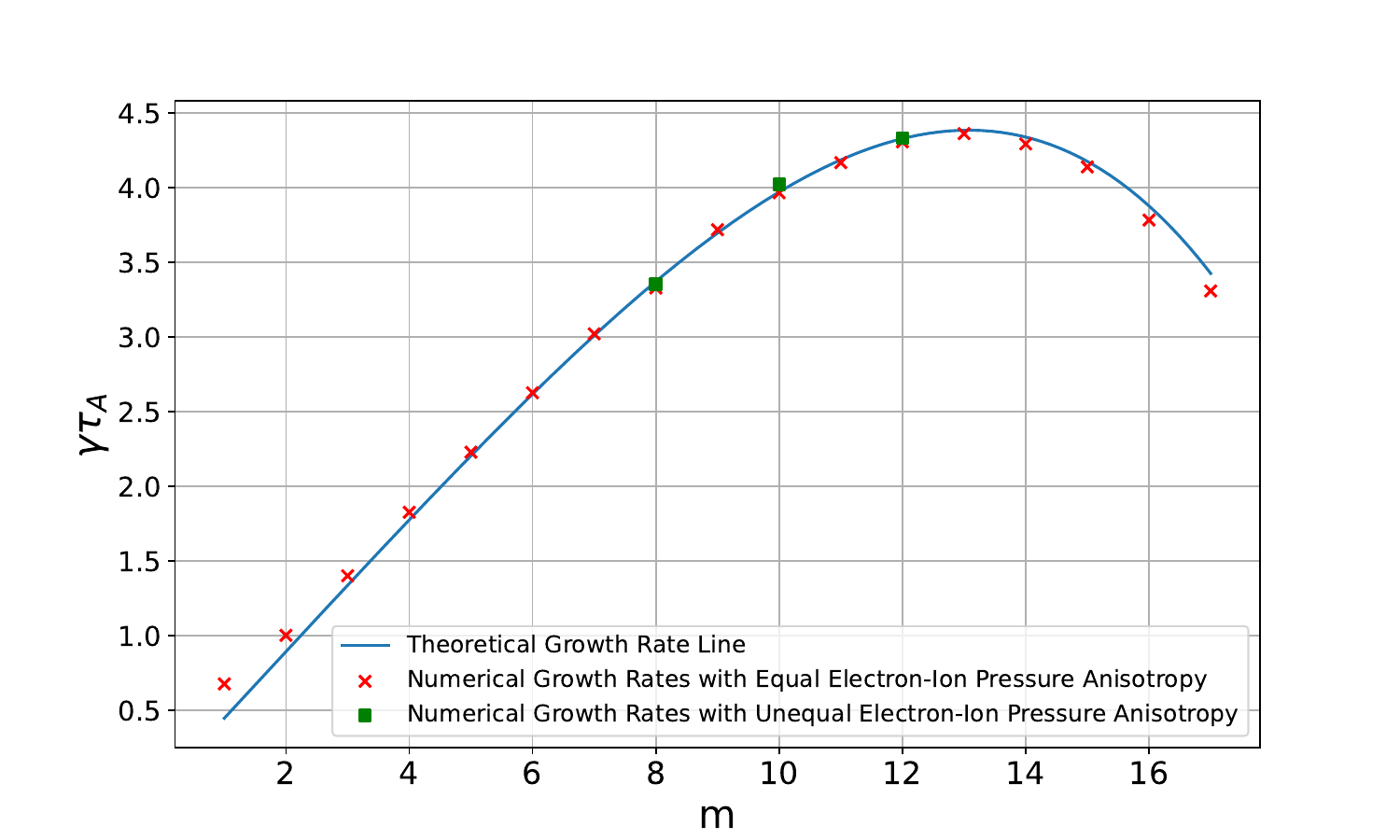}
\caption{Normalized growth rate of the firehose mode, $\gamma \tau_A$, versus mode number, $m=2\pi kL$, over a span of unstable $m$ values. Red crosses show values from simulations with equal electron and ion pressure anisotropies;  values for unequal anisotropies are shown as green squares. The blue line is the theoretical growth rate.
\label{fig:fire}}
\end{figure}

In the second benchmark, the rate of growth of the firehose instability was explored with a pre-set pressure anisotropy above the instability threshold ($\alpha^2 = -0.2$). The system was initialized with small amplitude periodic disturbances with seventeen distinct wavelengths, defined by $k=m/2\pi L$, where $m$ is the mode number. The simulations incorporated a kinematic viscosity of $\nu = 5.0 \times 10^{-5}$. The growth rate, which includes the pressure anisotropy drive and viscous damping is given by $\gamma = k C_A | \alpha | - \nu k^4$. The fourth-order hyperviscosity sets the upper bound of instability at small spatial scales. Figure \ref{fig:fire} presents the expected theoretical growth rate (represented by a solid blue line) and the growth rates found in the simulations for a range of unstable modes.  The red crosses indicate systems in which the electrons and ions contribute equally to the total anisotropy. To explore the effect of unequal pressure anisotropies, we also performed several simulations where the ions contribute 3/4 and the electrons contribute 1/4 of the total anisotropy (green squares). The growth rates measured in the unequal anisotropy simulations closely follow those from equal anisotropy simulations and the results from all the simulations are consistent with the predictions of linear theory. Note that there are no spurious unstable modes observed at short scales such as those that arise in some numerical models when the mode frequency exceeds the cyclotron frequency \citep{Tronci14}. All waves in the {\it kglobal} model are below the cyclotron frequency. 

\section{Conclusions and Discussion} \label{sec:conclusion}
The \textit{kglobal} model has been developed to explore magnetic-reconnection-driven particle acceleration in macroscale systems. The basis for the model is that particle energization is controlled by the dynamics of reconnection-driven flux ropes and turbulence at the macroscale rather than in kinetic-scale boundary layers. In the present paper we develop equations that extend the \textit{kglobal} to include the dynamics of particle ions. This requires the inclusion of the inertia of the ion particles into the fluid momentum equation. The model correctly captures the development of the pressure anisotropy on the dynamics of the magnetic field, which has a significant influence on reconnection dynamics. 

Our future work will center on using the upgraded \textit{kglobal} model to investigate the relative acceleration and energization of electrons and ions during magnetic reconnection in macroscale systems. Fundamental questions will be addressed by this model.  Can reconnection produce extended power-law distributions of energetic ions and electrons?  What is the relative partition of energy into nonthermal ions and electrons? Can reconnection produce ion and electron energy spectra consistent with {\it in situ} measurements from NASA's Magnetospheric Multiscale Mission in the Earth's magnetotail and from the Parker Solar Probe in the near solar environment?


\begin{acknowledgements}
We acknowledge support from the FIELDS team of the Parker Solar Probe (NASA Contract No. NNN06AA01C), the NASA Drive Science Center on Solar Flare Energy Release (SolFER) under Grant 80NSSC20K0627, NASA Grants 80NSSC20K1277 and 80NSSC22K0352 and NSF Grant PHY2109083. The simulations were carried out at the
National Energy Research Scientific Computing Center (NERSC).
The data used to perform the analysis and construct the figures for this paper are preserved at the NERSC High  Performance Storage System and are available upon request.
\end{acknowledgements}

%






\appendix

\section{Derivation of the Fluid Ion Momentum Equation  }\label{sec:ap_der}

The derivation of the fluid ion momentum equation is complicated by the need to keep the inertia of the particle ions.  We consider four distinct species, fluid ions, fluid electrons, particle electrons, and particle ions with number densities $n_{if}$, $n_{ef}$, $n_{ep}$, and $n_{ip}$, respectively, that satisfy the charge neutrality constraint 
\begin{equation}\label{equ:neu1}
n_{i f}+n_{i p}-n_{e f}-n_{e p}=0
\end{equation}
where $n_{if}$ is advanced in time with a continuity equation. The motion of the particle species includes the $\mathbf{E}\times\mathbf{B}$ perpendicular to $\mathbf{B}$ and parallel dynamics at the relevant local velocity along  $\mathbf{B}$. The number densities $n_{ip}$ and $n_{ep}$ are computed on the grid so both particle species also effectively satisfy continuity equations. The electron fluid also satisfies a continuity equation because of charge neutrality and because the three other species satisfy continuity equations. Specifically  Eq.~(\ref{equ:neu1}) for charge neutrality is used to  determine $n_{ef}$. In addition, each species satisfies a momentum equation:
\begin{equation}\label{equ:ap_if_mom1}
\frac{\partial\left(m_{i} n_{i f} \mathbf{v}_{i f}\right)}{\partial t}+\boldsymbol{\nabla} \cdot\left(m_{i} n_{i f} \mathbf{v}_{i f} \mathbf{v}_{i f}\right)=e n_{i f} \mathbf{E}+\mathbf{J}_{i f} \times \mathbf{B} / c-\nabla P_{i f}
\end{equation}
for fluid ions, 
\begin{equation}\label{equ:ap_ef_mom1}
\frac{\partial\left(m_{e} n_{e f} \mathbf{v}_{e f}\right)}{\partial t}+\boldsymbol{\nabla} \cdot\left(m_{e} n_{e f} \mathbf{v}_{e f} \mathbf{v}_{e f}\right)=-e n_{e f} \mathbf{E}+\mathbf{J}_{e f} \times \mathbf{B} / c-\nabla P_{e f}
\end{equation}
for fluid electrons, 
\begin{equation}\label{equ:ap_ep_mom1}
\frac{\partial\left(m_{e} n_{e p} \mathbf{v}_{e p}\right)}{\partial t}=-e n_{e p} \mathbf{E}+\mathbf{J}_{e p} \times \mathbf{B} / c-\boldsymbol{\nabla} \cdot \mathbb{T}_{e p}
\end{equation}
for particle electrons, and 
\begin{equation}\label{equ:ap_ip_mom1}
\frac{\partial\left(m_{i} n_{i p} \mathbf{v}_{i p}\right)}{\partial t}=e n_{i p} \mathbf{E}+\mathbf{J}_{i p} \times \mathbf{B} / c-\boldsymbol{\nabla} \cdot \mathbb{T}_{i p}
\end{equation}
for particle ions.  In these equations:
\begin{itemize}
    \item \( m_{i} \) and \( m_{e} \): Represent the masses of ions and electrons, respectively.
    \item \( \mathbf{v}_{i p} \), \( \mathbf{v}_{e f} \), \( \mathbf{v}_{e p} \) and \( \mathbf{v}_{i f} \): Denote the velocities of fluid ions, fluid electrons, particle electrons, and particle ions, respectively.
    \item \( \mathbf{J}_{i f} \), \( \mathbf{J}_{e f} \), \( \mathbf{J}_{e p} \), and \( \mathbf{J}_{i f} \): Represent the current densities corresponding to fluid ions, fluid electrons,  particle electrons, and particle ions, respectively.
    \item \( P_{i f} \) and \( P_{e f} \): Denote the (scalar) pressures for fluid ions and fluid electrons.
    \item \( \mathbb{T}_{e p} \) and \( \mathbb{T}_{i p} \): Represent the stress tensor for particle electrons and particle ions.
\end{itemize}
The choice to use the stress tensor in the particle equations, rather than the mathematically equivalent pressure and convective terms in the fluid equations, simplifies the remainder of the derivation. We emphasize that the electron equations are unchanged from those in the original {\it kglobal} model. As in the original model
all species are assumed to move perpendicular to the magnetic field with the $\mathbf{E} \times \mathbf{B}$ drift velocity.  As discussed in Sec.~\ref{subsec:kglobal}, the parallel velocity of the electrons is ordered to be much larger than the perpendicular velocity so that the parallel inertia of electros is retained while the perpendicular inertia is discarded:
\begin{equation}\label{eq:ve_relation}
v_{e f, \|} \sim v_{e p, \|} \gg v_{e p, \perp}\sim v_{e f, \perp}.
\end{equation}
With these approximations equation \eqref{equ:ap_ef_mom1}, reduces to   
\begin{equation}\label{eq:qp_final_equation}
\frac{\partial\left(m_{e} n_{e f} v_{e f \|} \mathbf{b}\right)}{\partial t}=-e n_{e f} \mathbf{E}+\mathbf{J}_{e f} \times \mathbf{B} / c-\nabla P_{e f}-\mathbf{b} \mathbf{B} \cdot \nabla \frac{m_{e} n_{e f} v_{e f, \|}^{2}}{B}-m_{e} n_{e f} v_{e f, \|}^{2} \boldsymbol{\kappa},
\end{equation}
where $\mathbf{b}$ is a unit vector parallel to the local magnetic field and  $\boldsymbol{\kappa} = \mathbf{b} \cdot \boldsymbol{\nabla} \mathbf{b}$ is the curvature vector.
Similarly, the parallel particle electron velocity becomes
\begin{equation}\label{eq:ap_fluid_electrons2}
\frac{\partial\left(m_{e} n_{e p} v_{e p, \|} \mathbf{b}\right)}{\partial t}=-e n_{e p} \mathbf{E}+\mathbf{J}_{ep} \times \mathbf{B} / c-\boldsymbol{\nabla\cdot} \mathbb{T}_{e p}.
\end{equation}
Again, Eqs.~(\ref{eq:qp_final_equation}) and (\ref{eq:ap_fluid_electrons2}) are unchanged from the original {\it kglobal} model. Summing equations \eqref{equ:ap_if_mom1},  \eqref{equ:ap_ip_mom1}, \eqref{eq:qp_final_equation} and \eqref{eq:ap_fluid_electrons2}, using the quasi-neutrality of equation \eqref{equ:neu1}, which eliminates \textbf{E}, we obtain
\begin{equation}
\begin{aligned}\label{equ:ap_comb}
& m_{i} \frac{\partial\left(n_{i} \mathbf{v}_{i f}\right)}{\partial t}+m_{i} \frac{\partial\left(n_{i p}\left(v_{i p,\parallel}-v_{i f,\parallel}\right)\mathbf{b}\right)}{\partial t}
+\frac{m_{i}}{e} \boldsymbol{\nabla} \cdot\left(\mathbf{J}_{i f} \mathbf{v}_{i f}\right) \\
& = \mathbf{J} \times \mathbf{B} / c-\boldsymbol{\nabla} P_{i f}-\boldsymbol{\nabla} P_{e f}-\mathbf{b} \mathbf{B} \cdot \boldsymbol{\nabla} \frac{m_{e} n_{e f} v_{e f, \|}^{2}}{B} -m_{e} n_{e f} v_{e f, \|}^{2} \boldsymbol{\kappa}-\boldsymbol{\nabla \cdot} (\mathbb{T}_{e p}+\mathbb{T}_{ip})
\end{aligned}
\end{equation}
where $n_i = n_{ip} + n_{if}$ is the total ion density.  Terms proportional to the electron mass on the left-hand side (LHS) of Equation \eqref{equ:ap_comb} have been discarded as discussed in Sec.~\ref{subsec:kglobal}. 

The derivation of the large-scale parallel electric field, $E_\|$, which is important in calculating the motion of particle electrons (see Eq.~(\ref{eq:ep_par})), follows exactly as in \cite{Arnold19}.  The parallel components of the four momentum equations are summed, the time rate of change of the parallel current is discarded, and we obtain an expression for $E_\|$,
\begin{equation}
e n_{i} E_{\|}=-\mathbf{b} \cdot \nabla \cdot \mathbb{T}_{e p}-\mathbf{b} \cdot \nabla P_{e f}-\mathbf{B} \cdot \nabla \frac{m_{e} n_{e f} v_{e f, \|}^{2}}{B}
\end{equation}
Note that all of the ion contributions have dropped out and that the result is unchanged from that in Equation \ref{equ:e_par}, except that $n_i = n_{ip} + n_{if}$.
 
Now the expression for the second term on the LHS of equation \eqref{equ:ap_comb} can be transformed as follows:

\begin{equation}\label{equ:ap_process2}
\begin{aligned}
& \frac{\partial }{\partial t} n_{i p}\left(v_{i p\parallel}-v_{i f\parallel}\right)\mathbf{b} = \mathbf{b}\left(v_{i p \|}-v_{i f \|}\right) \frac{\partial n_{i p}}{\partial t}+n_{i p} \frac{v_{i p \|}-v_{i f \|}}{B} \frac{\partial \mathbf{B}}{\partial t} +\mathbf{B} n_{i p} \frac{\partial}{\partial t}\left(\frac{v_{i p \|}-v_{i f \|}}{B}\right)
\end{aligned}
\end{equation}
Using the parallel component of the momentum equation of fluid ions in Equation \eqref{equ:ap_if_mom1}, the continuity equation, and the chain rule,
we obtain the expression:
\begin{equation}\label{equ:ap_process5}
\begin{aligned}
\frac{\partial}{\partial t} n_{i p}\left(v_{i p\parallel}-v_{i f\parallel}\right)\mathbf{b} & = \mathbf{b}\left(v_{i p \|}-v_{i f \|}\right) \frac{\partial n_{i p}}{\partial t} + n_{i p} \frac{v_{i p \|}-v_{i f \|}}{B}(\mathbb{I}-\mathbf{b b}) \cdot \frac{\partial \mathbf{B}}{\partial t} \\
&\quad + n_{i p} \mathbf{b}\left( -\mathbf{b} \cdot\left(\mathbf{v}_{i p} \cdot \nabla \mathbf{v}_{i p}\right) + \mathbf{b} \cdot\left(\mathbf{v}_{i f} \cdot \nabla \mathbf{v}_{i f}\right) - \frac{\mathbf{b} \cdot \boldsymbol{\nabla} \cdot \mathbb{P}_{i p}}{m_{i} n_{i p}} + \frac{\mathbf{b} \cdot \boldsymbol{\nabla} P_{i f}}{m_{i} n_{i f}} \right)
\end{aligned}
\end{equation}
The stress tensor $\mathbb{T}$ and pressure tensor $\mathbb{P}$ can be linked via
\begin{equation}\label{equ:stress_pressure_t}
\mathbf{b b} \cdot \boldsymbol{\nabla} \cdot \mathbb{T} = \mathbf{b b} \cdot \boldsymbol{\nabla} \cdot \mathbb{P} + \mathbf{b b} \cdot \boldsymbol{\nabla} \cdot m n \mathbf{v} \mathbf{v} = \mathbf{b b} \cdot \boldsymbol{\nabla} \cdot \mathbb{P} + \mathbf{b} \mathbf{b} \cdot\left( m n \mathbf{v} \cdot \boldsymbol{\nabla} \mathbf{v} - m \mathbf{v} \frac{\partial n}{\partial t} \right)
\end{equation}
From the combination of Equations \eqref{equ:ap_comb}, \eqref{equ:e_par}, and \eqref{equ:ap_process5}, plus some algebraic manipulations
we can obtain the following equivalent expressions:
\begin{equation}
\begin{split}
\frac{m_i}{e}\frac{\partial \mathbf{J}_{if}}{\partial t} = &-\frac{m_i}{e}\boldsymbol{\nabla\cdot}(\mathbf{J}_{if}\mathbf{v}_{if}) - \boldsymbol{\nabla}P_{if} + \frac{n_{if}}{n_i}\bigg(\mathbf{J}\boldsymbol{\times}\mathbf{B}/c - \boldsymbol{\nabla}_{\perp}P_{ef} - (\boldsymbol{\nabla\cdot}\mathbb{T}_{ep})_{\perp} - m_en_{ef}v^2_{\parallel,ef}\boldsymbol{\kappa} + en_iE_{\parallel}\mathbf{b}\bigg) \\
&+ (\mathbb{I}-\mathbf{b}\mathbf{b})\boldsymbol{\cdot}\bigg[\frac{n_{ip}}{n_i}\left(\boldsymbol{\nabla}P_{if} + \frac{m_e}{e}(\mathbf{J}_{if}\boldsymbol{\cdot\nabla})\mathbf{v}_{if}\right) -\frac{n_{if}}{n_i}\left(\boldsymbol{\nabla\cdot}\mathbb{P}_{ip} + \frac{m_e}{e}(\mathbf{J}_{ip}\boldsymbol{\cdot\nabla})\mathbf{v}_{ip}\right) -m_i\frac{n_{ip}n_{if}}{n_i}\frac{v_{\parallel,ip}-v_{\parallel,if}}{B}\frac{\partial \mathbf{B}}{\partial t}\bigg]
\end{split}
\end{equation}
and
\begin{equation}
\begin{split}
m_i\frac{\partial\, (n_{i}\mathbf{v}_{if})}{\partial t} &=- m_i\boldsymbol{\nabla \cdot} (n_{if}\mathbf{v}_{if}\mathbf{v}_{if}) - \boldsymbol{\nabla}P_{if} +
\mathbf{J}\boldsymbol{\times}\mathbf{B}/c - \boldsymbol{\nabla}_{\perp}P_{ef}- (\boldsymbol{\nabla\cdot}\mathbb{T}_{ep})_{\perp}- m_en_{ef}v^2_{ef,\parallel}\boldsymbol{\kappa}+en_iE_{\parallel}\mathbf{b} \\&+ 
\mathbf{b}\,m_iv_{if\parallel}\bigg(\frac{n_{ip}}{n_{if}}\boldsymbol{\nabla\cdot}n_{if}
v_{if\parallel}\mathbf{b}- \boldsymbol{\nabla\cdot}n_{ip}
v_{ip\parallel}\mathbf{b}-n_{if}\mathbf{v}_{i,\perp}\boldsymbol{\cdot\nabla}
\frac{n_{ip}}{n_{if}}\bigg) - m_in_{ip}\frac{v_{ip\parallel}-v_{if\parallel}}{B} (\mathbb{I}-\mathbf{b}\mathbf{b}) \boldsymbol{\cdot} \frac{\partial \mathbf{B}}{\partial t} \\ &-\frac{n_{ip}}{n_{if}}\mathbf{b}\mathbf{b}\boldsymbol{\cdot\nabla\cdot}\mathbb{T}_{if} -
(\boldsymbol{\nabla\cdot}\mathbb{T}_{ip})_{\perp}
\end{split}
\end{equation}
the second of which is Equation \eqref{eq:if_mom2} of Section \ref{sec:model}.

\section{Energy conservation}
Energy conservation is most simply explored using the equations of motion for each of the four species in Equations \eqref{equ:ap_if_mom1}-\eqref{equ:ap_ip_mom1} rather than Equation \eqref{eq:if_mom2}. Starting from these more primitive equations is valid because nothing is discarded in the manipulation of the ion equations and in the electron equations the only discarded terms are the inertia associated with the perpendicular motion and this was previously shown to lead a set of equations that conserve energy \citep{Arnold19}. Each of the equations has the generic form 
\begin{equation}\label{equ:ap_mom_j}
\frac{\partial\left(m_jn_j \mathbf{v}_j\right)}{\partial t}+\boldsymbol{\nabla}\cdot\left(m_jn_j\mathbf{v}_j\mathbf{v}_j\right)=q_j n_j\mathbf{E}+\mathbf{J}_j\times \mathbf{B} / c-\boldsymbol{\nabla} \cdot \mathbb{P}_j,
\end{equation}
where the subscript $j$ denotes the jth species. Taking the dot product with $\mathbf{v}_j$ and using the continuity equation, we obtain
\begin{equation}\label{equ:ap_energy_j}
\frac{\partial}{\partial t}\left( m_jn_jv_j^2/2\right)+\boldsymbol{\nabla}\cdot\left(m_jn_jv_j^2\mathbf{v}_j/2\right)=\mathbf{J}_j\cdot\mathbf{E}-\mathbf{v}_j\cdot\left(\boldsymbol{\nabla} \cdot \mathbb{P}_j\right).
\end{equation}
For a generic gyrotropic pressure tensor $\mathbb{P}_j$, Equation \eqref{equ:ap_energy_j} takes the form
\begin{equation}\label{equ:ap_W_flux}
    \frac{\partial W_j}{\partial t}+\boldsymbol{\nabla}\cdot\mathbf{Q}_j=\mathbf{J}_j\cdot\mathbf{E},
\end{equation}
where 
\begin{equation}\label{equ:ap_W}
    W=\frac{1}{2}m_jn_jv_j^2+\frac{1}{2}P_\|+P_\perp,
\end{equation}
is the energy density and
\begin{equation}
    \mathbf{Q}=\frac{1}{2}m_jn_jv_j^2\mathbf{v}_j+\left(\frac{3}{2}P_{j\|}+P_{j\perp}\right)v_{j\|}\mathbf{b}+\left(\frac{1}{2}P_{j\|}+2P_{j\perp}\right)\mathbf{v}_{j\perp}
\end{equation}
is the energy flux. For electrons only the bulk flow kinetic energy associated with the parallel motion needs to be retained. The bulk flow kinetic energy associated with the electron perpendicular motion can be neglected. Finally, Equation (\ref{equ:ap_W_flux}) for the total energy of the particle species also follows directly from the guiding center equation for the parallel momentum and $\mu$ conservation. Summing the energy equations of the four species and integrating over all space yields the energy equation
\begin{equation}\label{equ:ap_energy_total}
\int\mathbf{dx}\frac{\partial}{\partial t}\left(W_{if}+W_{ip}+W_{ef}+W_{ep}\right)=\int\mathbf{dx}\mathbf{J}\cdot\mathbf{E}
\end{equation}
where $\mathbf{J}=\mathbf{J}_{if}+\mathbf{J}_{ip}+\mathbf{J}_{ef}+\mathbf{J}_{ep}$ is the total current. From Faraday's law
\begin{equation}\label{equ:ap_faraday}
    \frac{\partial}{\partial t}\frac{B^2}{8\pi}+\frac{c}{4\pi}\boldsymbol{\nabla}\cdot\left(\mathbf{E}\times\mathbf{B}\right)+\mathbf{J}\cdot\mathbf{E}=0.
\end{equation}
The final energy conservation law then takes the form
\begin{equation}\label{equ:ap_energy_int}
    \int\mathbf{dx}\frac{\partial}{\partial t}\left(W_{if}+W_{ip}+W_{ef}+W_{ep}+B^2/8\pi\right)=0.
\end{equation}
The derivation of energy conservation from the four individual momentum equations (see Eq.~(\ref{equ:ap_mom_j})) is exact. However, in the total fluid ion momentum equation in Eq.~(\ref{eq:if_mom2}), the perpendicular momenta in the electron fluid and particle stress tensors and comparable time-dependent inertial terms have been discarded. The discarded terms scale like $m_e/m_i$ compared with those retained. Also, the parallel electric field in Faraday's law (Eq.~(\ref{eq:b})) is discarded compared with $\mathbf{E}_\perp$ since their ratio scales like $\rho_s/L$ with $\rho_s$ the ion sound Larmor radius. The consequence is that energy conservation in {\it kglobal} is not exact but violations are small. 

An alternative derivation of energy conservation that parallels that in the original {\it kglobal} formulation \citep{Arnold19}, can be obtained by defining a center-of-mass (CM) velocity $n_i\mathbf{v}_{icm}=n_{ip}\mathbf{v}_{ip}+n_{if}\mathbf{v}_{if}$. The equation for the CM velocity is obtained by summing the two ion momentum equations (Eqs.~(\ref{equ:ap_if_mom1}) and (\ref{equ:ap_ip_mom1})),
\begin{equation}\label{eq:ap_cm_mom}
\begin{split}
&\frac{\partial}{\partial t} (m_in_{i}\mathbf{v}_{icm})  +\boldsymbol{\nabla}\cdot(m_in_i\mathbf{v}_{icm}\mathbf{v}_{icm})=en_i\mathbf{E}+\frac{1}{c}n_ie\mathbf{v}_{icm}\times\mathbf{B}-\boldsymbol{\nabla \cdot}\mathbb{P}_{icm},
\end{split}
\end{equation}
where the CM pressure tensor is given by
\begin{equation}
  \mathbb{P}_{icm} =m_i[n_{ip}(\mathbf{v}_{ip}-\mathbf{v}_{icm})(\mathbf{v}_{ip}-\mathbf{v}_{icm})+(n_{if}(\mathbf{v}_{if}-\mathbf{v}_{icm})(\mathbf{v}_{if}-\mathbf{v}_{icm})] +\mathbb{P}_{ip}+\mathbb{I}P_{if},
\end{equation}
where the velocity dependent terms produce an effective presssure associated with the differences between the two ion velocities and the CM velocity. This equation for the CM velocity is then summed with the two electron momentum equations (Eqs.~(\ref{equ:ap_ef_mom1}) and (\ref{equ:ap_ep_mom1})) to produce
\begin{equation}\label{eq:ap_cm_mom1}
\begin{split}
&\frac{\partial}{\partial t} (m_in_{i}\mathbf{v}_{icm})  +\boldsymbol{\nabla}\cdot(m_in_i\mathbf{v}_{icm}\mathbf{v}_{icm})=\frac{1}{c}\mathbf{J}\times\mathbf{B}-\boldsymbol{\nabla \cdot}\mathbb{P}_{icm} -(\boldsymbol{\nabla \cdot}\mathbb{T}_{ef})_\perp -
(\boldsymbol{\nabla\cdot}\mathbb{T}_{ep})_{\perp}+en_iE_{\parallel}\mathbf{b},
\end{split}
\end{equation}
The energy equation in CM coordinates then follows by taking the dot product of this equation with $\mathbf{v}_{icm}$,
\begin{equation}\label{eq:ap_cm_energy}
\begin{split}
&\frac{\partial}{\partial t} (\frac{1}{2}m_in_{i}v_{icm}^2)  +\boldsymbol{\nabla}\cdot(\frac{1}{2}m_in_iv_{icm}^2\mathbf{v}_{icm})=\mathbf{J}_\perp\cdot\mathbf{E}_\perp-\mathbf{v}_{icm}\cdot\boldsymbol{\nabla \cdot}\mathbb{P}_{icm}\\& -\mathbf{v}_{icm\perp}\cdot[(\boldsymbol{\nabla \cdot}\mathbb{T}_{ef})_\perp -
(\boldsymbol{\nabla\cdot}\mathbb{T}_{ep})_{\perp}]+en_iv_{icm\parallel}E_{\parallel}.
\end{split}
\end{equation}
The contribution from the CM pressure tensor is linked to the ion thermal energy in the CM frame, $W_{icmP}$, as in an MHD system. $W_{icmP}$ is given by
\begin{equation}
\begin{split}
    W_{icmP} &=\frac{1}{2}m_i[n_{ip}(\mathbf{v}_{ip}-\mathbf{v}_{icm})^2+n_{if}(\mathbf{v}_{if}-\mathbf{v}_{icm})^2] +\frac{1}{2}P_{ip\parallel}+P_{ip\perp}+\frac{3}{2}P_{if}\\&
    =
  \frac{1}{2}m_i[n_{ip}\mathbf{v}_{ip}^2+n_{if}\mathbf{v}_{if}^2-n_i\mathbf{v}_{icm}^2] +\frac{1}{2}P_{ip\parallel}+P_{ip\perp}+\frac{3}{2}P_{if}.
\end{split}
\end{equation}
The time derivative of $W_{icmP}$ then follows from the time derivative of the individual kinetic energies as in Eq.~(\ref{equ:ap_energy_j}),
\begin{equation}\label{equ:ap_energy}
    \frac{\partial W_{icmP}}{\partial t}+\boldsymbol{\nabla}\cdot(\mathbf{Q}_{ip}+\mathbf{Q}_{if})-\boldsymbol{\nabla}\cdot(\frac{1}{2}n_iv_{icm}^2\mathbf{v}_{icm})=-\mathbf{v}_{icm}\cdot\mathbb{P}_{icm}.
\end{equation}
This equation is then combined with Eq.~(\ref{eq:ap_cm_energy}) to eliminate the $\mathbb{P}_{icm}$
term,
\begin{equation}\label{eq:ap_cm_energy1}
\begin{split}
&\frac{\partial}{\partial t} (\frac{1}{2}m_in_{i}v_{icm}^2+W_{icmP})  +\boldsymbol{\nabla}\cdot(\mathbf{Q}_{ip}+\mathbf{Q}_{if})=\mathbf{J}\cdot\mathbf{E}-\mathbf{J}_{ef}\cdot\mathbf{E}-\mathbf{J}_{ep}\cdot\mathbf{E},
\end{split}
\end{equation}
where we have substituted $n_iev_{icm\parallel}=J_\parallel+n_{ef}ev_{ef\parallel}+n_{ep}ev_{ep\parallel}$, and used $\mathbf{v}_{icm\perp}=\mathbf{v}_{ep\perp}=\mathbf{v}_{ef\perp}$ since the Hall terms in Ohm's law have been ordered out. The contributions from $\mathbf{v}_{ef\perp}\cdot(\mathbb{T}_{ef})_\perp$ and $\mathbf{v}_{ep\perp}\cdot(\mathbb{T}_{ep})_\perp$ were calculated from their respective momentum equations with their perpendicular bulk flow inertia discarded. Since
\begin{equation}
    \frac{1}{2}m_in_iv_{icm}^2+W_{icmP}=W_{ip}+W_{if},
\end{equation}
the integral of Eq.~(\ref{eq:ap_cm_energy1}) yields Eq.~(\ref{equ:ap_energy_total}) and the final energy conservation law in Eq.~(\ref{equ:ap_energy_int}) follows as calculated previously.

\bibliography{kglobal_eqn}{}
\bibliographystyle{aasjournal}

\end{CJK*}
\end{document}